\documentclass[12pt]{article}
\usepackage{epsfig}
\usepackage{amsmath}
\usepackage{amssymb}
\newlength{\dinwidth}
\newlength{\dinmargin}
\begin{document}

%--------------------------------------------------------------------------
%--------------------------------------------------------------------------

\thispagestyle{empty}\rightline{Napoli DSF-T-16/2007}%
\rightline{INFN-NA-16/2007} \vspace*{2cm}

\begin{center}
{\LARGE Fully frustrated Josephson junction ladders with Mobius boundary
conditions as topologically protected qubits}

{\LARGE \ }

{\large Gerardo Cristofano\footnote{{\large {\footnotesize Dipartimento di
Scienze Fisiche,}{\it \ {\footnotesize Universit\'{a} di Napoli ``Federico
II''\ \newline
and INFN, Sezione di Napoli},}{\small Via Cintia, Compl.\ universitario M.
Sant'Angelo, 80126 Napoli, Italy}}}, Vincenzo Marotta\footnote{{\large
{\footnotesize Dipartimento di Scienze Fisiche,}{\it \ {\footnotesize %
Universit\'{a} di Napoli ``Federico II''\ \newline
and INFN, Sezione di Napoli},}{\small Via Cintia, Compl.\ universitario M.
Sant'Angelo, 80126 Napoli, Italy}}},} {\large Adele Naddeo\footnote{{\large
{\footnotesize Dipartimento di Fisica {\it ''}E. R. Caianiello'',}{\it \
{\footnotesize Universit\'{a} degli Studi di Salerno \ \newline
and CNISM, Unit\`{a} di Ricerca di Salerno, }}{\small Via Salvador Allende,
84081 Baronissi (SA), Italy}}}, Giuliano Niccoli\footnote{{\large
{\footnotesize LPTM, Universit\'{e} de Cergy-Pontoise, 2 avenue Adolphe
Chauvin, 95302 Cergy-Pontoise, France.}}}}

{\small \ }

{\bf Abstract\\[0pt]
}
\end{center}

\begin{quotation}
We show how to realize a ``protected''\ qubit by using a fully frustrated
Josephson Junction ladder (JJL) with Mobius boundary conditions. Such a
system has been recently studied within a twisted conformal field theory
(CFT) approach \cite{cgm2,cgm4} and shown to develop the phenomenon of flux
fractionalization \cite{noi4}. The relevance of a ``closed''\ geometry has
been fully exploited in relating the topological properties of the ground
state of the system to the presence of half flux quanta and the emergence of
a topological order has been predicted \cite{noi3}. In this letter the
stability and transformation properties of the ground states under adiabatic
magnetic flux change are analyzed and the deep consequences on the
realization of a solid state qubit, protected from decoherence, are
presented.

\vspace*{0.5cm}


{\footnotesize PACS: 11.25.Hf, 03.75.Lm, 74.81.Fa\newpage }\baselineskip%
=18pt \setcounter{page}{2}
\end{quotation}

Arrays of weakly coupled Josephson junctions provide an experimental
realization of the two dimensional ($2D$) XY model. A Josephson junction
ladder (JJL) is the simplest quasi-one dimensional version of an array in a
magnetic field \cite{ladder}; recently such a system has been the subject of
many investigations because of its possibility to display different
transitions as a function of the magnetic field, temperature, disorder,
quantum fluctuations and dissipation. In a recent paper \cite{noi4} we
analyzed the phenomenon of fractionalization of the flux quantum $\frac{hc}{%
2e}$ in a fully frustrated JJL in order to investigate how the phenomenon of
Cooper pair condensation could cope with properties of charge (flux)
fractionalization, typical of a low dimensional system with a discrete $Z_{2}
$ symmetry. The role of such a symmetry was recognized to be crucial for
demanding more general boundary conditions, of the Mobius type, at the end
sites of the ladder \cite{noi4}. The same feature was evidenced also in
quantum Hall systems in the presence of impurities or defects \cite{noi1}
\cite{noi2}\cite{noi5}. Furthermore a $Z_{2}$ symmetry is present in the
fully frustrated XY (FFXY) model or equivalently, see Refs. \cite{foda}\cite
{noi}, in two dimensional Josephson junction arrays (JJA) with half flux
quantum $\frac{1}{2}\frac{hc}{2e}$ threading each square cell and accounts
for the degeneracy of the ground state. We noticed how it was possible to
generate non trivial topologies, i.e. the torus, in the context of a CFT
approach, which allowed us to construct a ground state wave function, whose
center of charge could describe a coherent superposition of localized states
sharing all the non trivial global properties of the order parameter. In
particular for the FFXY model they were shown to be closely related to the
presence of half flux quanta, also viewed as topological defects\cite{noi4}.
The emergence of topological order in fully frustrated JJLs with non trivial
geometry has been predicted and fully exploited in Ref. \cite{noi3} by means
of CFT techniques. Such a concept was first introduced in order to describe
the ground state of a quantum Hall fluid \cite{wen} but today it is of much
more general interest \cite{wen1}. Two features of topological order are
very striking: fractionally charged quasiparticles and a ground state
degeneracy depending on the topology of the underlying manifold, which is
lifted by quasiparticles tunneling processes. In general a system is in a
topological phase if its low-energy, long-distance effective field theory is
a topological quantum field theory that is, if all of its physical
correlation functions are topologically invariant up to corrections of the
form $e^{-\frac{\Delta }{T}}$at temperature $T$ for some nonzero energy gap $%
\Delta $.{\bf \ }More recently superconductors have been proposed in which
superconductivity arises from a topological mechanism rather than from a
Ginzburg-Landau paradigm: the key feature is a mapping on an effective
Chern-Simons gauge theory, which turns out to be exact in the case of JJA
and frustrated JJA \cite{pasquale}. As we will stress in the following,
topological order is crucial for the implementation of fully frustrated JJLs
as ``protected'' qubits \cite{ioffe}\cite{kitaev} in solid state quantum
computation realm. The idea in all such realizations is that the systems
involved (large and small size Josephson junction arrays of special geometry
\cite{ioffe}\cite{ioffe1}) share the property that, in the classical limit
for the local superconducting variables, the ground state is highly
degenerate. The residual quantum processes within such a low energy subspace
lift the classical degeneracy in favor of macroscopic coherent
superpositions of classical ground states \cite{ioffe1}. An example of such
a system has been proposed, which consists of chains of rhombi frustrated by
an half flux quantum \cite{ioffe1} with the property that in the classical
limit each rhombus has two degenerate states. The protected degeneracy in
all such systems emerges as a natural property of the lattice Chern-Simons
gauge theories which describe them \cite{ioffe1}. In general, if a physical
system has topological degrees of freedom that are insensitive to local
perturbations (that is noise), then information contained in those degrees
of freedom would be automatically protected against errors caused by local
interactions with the environment \cite{kitaev}.

The aim of this letter is to show how to realize a ``protected''\ qubit in
terms of a fully frustrated Josephson Junction ladder (JJL) with Mobius
boundary conditions by fully exploiting the implications of ``closed''\
geometries on the ground state global properties of the system, already
studied in Ref. \cite{noi4}. Such a qubit would be the elementary building
block of a ``protected'' quantum computer. The task appears to be not very
simple; in general we need a quantum system with $2^{K}$\ quantum states ($K$%
\ being the number of big openings in the Josephson system under study)
which are degenerate in the absence of external perturbations and are robust
against local random fluctuations, that is against noise. This means that
any coupling to the environment doesn't induce transitions between the $%
2^{K} $ quantum states or change their relative phases.
Summarizing, we need a system, whose Hilbert space contains a
$2^{K}$-dimensional subspace characterized by the crucial property
that any local operator $\widehat{O}$ has only state-independent
diagonal matrix elements up to vanishingly small corrections:
$\left\langle n\right| \widehat{O}\left| m\right\rangle
=O_{0}\delta _{mn}+o\left[ \exp \left( -L\right) \right] $, $L$
being the system size. A possible answer to such a highly non
trivial requirement could be a system with a protected subspace
built up by a topological degeneracy of the ground state
\cite{kitaev}. An alternative approach would be to exhibit a
low-energy effective field theory for the system under study which
is a topological one and whose vacua are topologically degenerate
and, then, robust against noise. This is the approach which we
follow in the present letter; in particular we show how to get a
protected subspace with $2^{K}$ quantum states, $K=1$, by
considering a Josephson junction ladder and closing it by imposing
Mobius boundary conditions, in order to get a non trivial
topology. We will show that such a system is described by a
low-energy effective field theory which is a twisted conformal
field theory \cite{cgm2,cgm4}\cite{noi1}\cite{noi2}. Such a theory
accounts very well for the topological properties of the system
under study \cite{noi4}\cite{noi3}. In particular we analyze the
stability and transformation properties of the ground state wave
functions under adiabatic magnetic flux change; in this way we are
able to identify the two states of a possible protected qubit and
also to describe its manipulation: ``flip state'' processes.

We recall that Josephson junction arrays (JJA) are a very useful tool for
investigating quantum-mechanical behaviour in a wide range of parameters
space, from $E_{C}\gg E_{J}$ (where $E_{C}=\frac{\left( 2e\right) ^{2}}{C}$
is the charging energy and $E_{J}=\frac{\hbar }{2e}I_{c}$ is the Josephson
coupling energy; $C$ is the capacitance of each island and $I_{c}$ is the
critical current of each junction) to $E_{J}$ $\gg E_{C}$. In fact there
exists a couple of conjugate quantum variables, the charge and phase of each
superconducting island, and two dual descriptions of the array can be given
\cite{fazio}: a) through the charges (Cooper pairs) hopping between the
islands, b) through the vortices hopping between the plaquettes. Furthermore
in the presence of an external magnetic field charges gain additional
Aharonov-Bohm phases and, conversely, vortices moving around islands gain
phases proportional to the average charges on the islands \cite{casher}.
Such basic quantum interference effects found applications in recent
proposals for solid state qubits for quantum computing, based on charge \cite
{hermon} or phase \cite{orlando1} degrees of freedom in JJAs. ``Charge''
devices operate in the regime $E_{C}\gg E_{J}$ while ``phase'' or ``flux''
devices are characterized by strongly coupled junctions with $E_{J}$ $\gg
E_{C}$.

Let us now focus on the simplest physical array one can devise in order to
meet all the above requests, that is a Josephson junction ladder with $N$
plaquettes closed in a ring geometry with a half flux quantum ($\frac{1}{2}%
\Phi _{0}=\frac{1}{2}\frac{hc}{2e}$) threading each plaquette \cite{ladder},
and describe briefly its general properties before introducing an
interaction of the charges (Cooper pairs) with a magnetic impurity (defect),
as drawn in Fig. 1. With each site $i$ we associate a phase $\varphi _{i}$
and a charge $q_{i}=2en_{i}$, representing a superconducting grain coupled
to its neighbors by Josephson couplings; $n_{i}$ and $\varphi _{i}$ are
conjugate variables satisfying the usual phase-number commutation relation.
The Hamiltonian describing the system is given by the quantum phase model
(QPM):
\begin{equation}
H=-\frac{E_{C}}{2}\sum_{i}\left( \frac{\partial }{\partial \varphi _{i}}%
\right) ^{2}-\sum_{\left\langle ij\right\rangle }E_{ij}\cos \left( \varphi
_{i}-\varphi _{j}-A_{ij}\right) ,  \label{act0}
\end{equation}
where $E_{C}=\frac{\left( 2e\right) ^{2}}{C}$ ($C$ being the capacitance) is
the charging energy at site $i$, while the second term is the Josephson
coupling energy between sites $i$ and $j$ and the sum is over nearest
neighbors. The most general form for the charging energy would be $\frac{1}{2%
}q_{i}C_{ij}^{-1}q_{j}$, where $C_{ij}^{-1}$\ is the inverse capacitance
matrix, but in this letter we assume for simplicity that the most important
contribution arises from the self-energy of each grain \cite{bradley}\cite
{ladder}. $A_{ij}=\frac{2\pi }{\Phi _{0}}$ $\int_{i}^{j}A{\cdot }dl$ is the
line integral of the vector potential associated to an external magnetic
field $B$ and $\Phi _{0}=\frac{hc}{2e}$ is the magnetic flux quantum. The
gauge invariant sum around a plaquette is $\sum_{p}A_{ij}=2\pi f$ with $f=%
\frac{\Phi }{\Phi _{0}}$, where $\Phi $ is the flux threading each plaquette
of the ladder. Let us label the phase fields on the two legs with $\varphi
_{i}^{\left( a\right) }$, $a=1,2$ and assume $E_{ij}=E_{x}$ for horizontal
links and $E_{ij}=E_{y}$ for vertical ones. Let us also make the gauge
choice $A_{ij}=+\pi f$ for the upper links, $A_{ij}=-\pi f$ for the lower
ones and $A_{ij}=0$ for the vertical ones, which corresponds to a vector
potential parallel to the ladder and taking opposite values on upper and
lower branches.

Thus the effective quantum Hamiltonian (\ref{act0}) can be written as \cite
{ladder}:
\begin{eqnarray}
-H &=&\frac{E_{C}}{2}\sum_{i}\left[ \left( \frac{\partial }{\partial X_{i}}%
\right) ^{2}+\left( \frac{\partial }{\partial \phi _{i}}\right) ^{2}\right] +
\nonumber \\
&&\sum_{i}\left[ 2E_{x}\cos \left( X_{i+1}-X_{i}\right) \cos \left( \phi
_{i+1}-\phi _{i}-\pi f\right) +E_{y}\cos \left( 2\phi _{i}\right) \right] ,
\label{ha2}
\end{eqnarray}
after performing the change of variables: $\varphi _{i}^{\left( 1\right)
}=X_{i}+\phi _{i}$, $\varphi _{i}^{\left( 2\right) }=X_{i}-\phi _{i}$, where
$X_{i}$, $\phi _{i}$ (i.e. $\varphi _{i}^{\left( 1\right) }$, $\varphi
_{i}^{\left( 2\right) }$) are only phase deviations of each order parameter
from the commensurate phase and should not be identified with the phases of
the superconducting grains \cite{ladder}.

When $f=\frac{1}{2}$ and $E_{C}=0$ (classical limit) the ground state of the
$1D$ frustrated quantum XY (FQXY) model displays - in addition to the
continuous $U(1)$ symmetry of the phase variables - a discrete $Z_{2}$
symmetry associated with an antiferromagnetic pattern of plaquette
chiralities $\chi _{p}=\pm 1$, measuring the two opposite directions of the
supercurrent circulating in each plaquette. Thus it has two symmetric,
energy degenerate, ground states characterized by currents circulating in
the opposite directions in alternating plaquettes in full analogy with the
checkerboard ground states of the $2D$ system \cite{teitel}. For small $%
E_{C} $ there is a gap for creation of kinks in the antiferromagnetic
pattern of $\chi _{p}$ and the ground state has quasi long range chiral
order \cite{ladder}; furthermore the charge noise, which is the strongest
noise, has less effect in such a regime \cite{ioffe}\cite{ioffe1}.

\begin{figure}[tbp]
\centering\includegraphics*[width=0.7\linewidth]{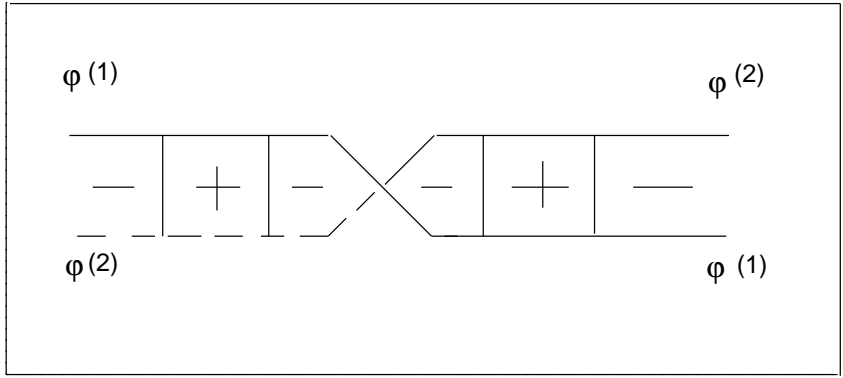}
\caption{JJL with a magnetic impurity }
\label{figura1}
\end{figure}

The ladder under study can be closed and arranged in a Corbino disk
geometry. As we will argue in the following, this is the relevant geometry
for the physical implementation of an ideal quantum computer. In closing the
ladder, we can distinguish two inequivalent configurations, corresponding to
an even or odd number of plaquettes in the ladder. It is due to the
antiferromagnetic pattern of plaquette chiralities, which characterizes the
JJL ground states.

- In the even case, the plaquettes on the opposite sides of the ladder have
opposite chiralities, for both the degenerate ground states. So, the closed
geometry can be realized gluing the opposite sides of the ladder keeping the
ground state antiferromagnetic pattern.

- In the odd case, the plaquettes on the opposite sides of the ladder have
the same chiralities, for both the degenerate ground states. In this case
the ladder has to be modified; a magnetic impurity has to be introduced, in
the glue-point, which couples the up and down phases through its interaction
with the Cooper pairs of the two legs (as represented in Fig.1).

In the odd configuration, the two degenerate ground states can be
represented by $\left| 0\right\rangle $ and $\left| 1\right\rangle $ and
distinguished by the value of the sum over all plaquettes $\sum_{p}\chi _{p}$
respectively equal to $-1$ and $+1$.

The outlined pattern, in this closed geometry, evidences the emerging of non
trivial topological properties intimately related to the twofold degeneracy
of the ground states which appear to be ``protected''\ from external
perturbations \cite{wen}\cite{noi3}\cite{noi4}. Moreover, such a pattern is
{\it size} independent, that is, it depends on the number of plaquettes only
by its party and, in particular, persists also in the continuum limit $%
N\rightarrow \infty $ and \ $a\rightarrow 0$, where $N$ is the number of
plaquettes, $a$ is the side-size of a plaquette and $L=aN$ is the constant
length of the ladder. \ Our strategy is to study the continuum limit of the
JJL in this closed geometry; indeed, in the continuum the powerful tools of
the CFT can be used to evince the topological properties of the system
which, when extended to finite ladder, allow us to propose a JJL realization
of a qubit device.

In the following, we will setup the CFT analysis, while some more details on
our twisted model (TM) can be found in the Appendix. Performing the
continuum limit of the Hamiltonian (\ref{ha2}), one obtains:
\begin{eqnarray}
-H &=&\frac{E_{C}}{2}\int dx\left[ \left( \frac{\partial }{\partial X}%
\right) ^{2}+\left( \frac{\partial }{\partial \phi }\right) ^{2}\right] +
\nonumber \\
&&\int dx\left[ E_{x}\left( \frac{\partial X}{\partial x}\right)
^{2}+E_{x}\left( \frac{\partial \phi }{\partial x}-\frac{\pi }{2}\right)
^{2}+E_{y}\cos \left( 2\phi \right) \right]  \label{ha3}
\end{eqnarray}
where we see that the $X$ and $\phi $ fields are decoupled. In fact the $X$
term of the above Hamiltonian is that of a free quantum field theory while
the $\phi $ one coincides with the quantum sine-Gordon model. Through an
imaginary-time path-integral formulation of such a model \cite{zinn} it can
be shown that the $1D$ quantum problem maps into a $2D$ classical
statistical mechanics system, the $2D$ fully frustrated XY model, where the
parameter $\alpha =\left( \frac{E_{x}}{E_{C}}\right) ^{\frac{1}{2}}$ plays
the role of an inverse temperature \cite{ladder}. We work in the regime $%
E_{x}\gg E_{y}$ where the ladder is well described by our TM with central
charge $c=2$.

Let us introduce in the continuum the closed geometry; in order to do so, we
require the compactification of the $\varphi ^{\left( a\right) }$ variables
to recover the angular nature of the up and down fields. Then the even and
odd configurations rising in the closed geometry of the finite ladder
correspond in the continuum to two different boundary conditions for the
fields, respectively, periodic ($P$) and Mobius ($A$) boundary conditions:
\begin{equation}
\varphi _{L}^{\left( 1\right) }\left( x=0\right) =+\varphi _{R}^{\left(
2\right) }\left( x=0\right) \text{ and }\varphi _{L}^{\left( 1\right)
}\left( x=0\right) =-\varphi _{R}^{\left( 2\right) }\left( x=0\right) ,
\label{blr}
\end{equation}
where we have indicated the compactified phases of the two legs as $\varphi
_{L}^{(1)}$and $\varphi _{R}^{(2)}$, $L$ and $R$ staying for left and right
components. Indeed in the limit of strong coupling the interaction between
the magnetic impurity at point $x=0$ (glue-point shown in Fig. 1) and the up
and down phases gives rise to these non trivial boundary conditions for the
fields \cite{noi1}. Such a Mobius condition is naturally satisfied by the
twisted field $\phi \left( z\right) $ of our TM (see eq. (\ref{phi}) in the
Appendix), which describes both the left moving component $\varphi
_{L}^{\left( 1\right) }$ and the right moving one $\varphi _{R}^{\left(
2\right) }$ in a folded description of a system with boundary \cite{noi1}
\cite{noi2}. In fact the TM results in a chiral description of the system
just described in terms of the chiral fields $X$ and $\phi $ (see eqs. (\ref
{X}), (\ref{phi})). The $m$-reduction technique \cite{cgm4} well accounts
for these non trivial boundary conditions for the JJ ladder due to the
presence of a topological defect, already built in the construction.

Our goal is the study of the stability and transformation properties of the
four ground states of the JJL in the closed geometry under an adiabatic
elementary flux change ($\pm \frac{hc}{2e}$) through the central hole of the
Corbino disk. Because of the energy gap, such an adiabatic transformation is
believed to leave the system in a ground state which can be different from
the original one, due to the occurrence of the ground state degeneracy. This
analysis will be the crucial step for the identification of the two states
of a possible protected qubit and for its manipulation: ``flip state''
processes.

We use the TM model to analyze such properties by standard conformal
techniques. In the CFT description the ground state wave functions are
expressed as correlation functions of the primary fields describing the
elementary particles, in our case the Cooper pairs. In particular in the
torus topology the characters of the theory are in one to one correspondence
with the ground states. Indeed, as we are going to show, they describe the
components of the ``center of charge'' for the corresponding ground state
wave functions \cite{gerardo}, which represent coherent states of Cooper
pairs on the torus. To such an extent, let us define for a single Cooper
pair on a torus $a\times b$ an effective mean-field Hamiltonian of the kind $%
H\left( x,y\right) =H_{0}\left( x,y\right) +V\left( x,y\right) $, where $%
H_{0}\left( x,y\right) =\left[ -i\hbar \overrightarrow{\nabla }-2e%
\overrightarrow{A}/c\right] ^{2}/2m$ is the Hamiltonian in the presence of
an uniform magnetic field and $V\left( x,y\right) $ is a mean-field scalar
potential such that $V\left( x,y\right) =V\left( x+a,y\right) =V\left(
x,y+b\right) $. It is now possible to define the magnetic translations
operators $\widetilde{{\cal S}}=e^{i\theta _{x}a/\hbar }$ and $\widetilde{%
{\cal T}}=e^{i\theta _{y}b/\hbar }$ along the two cycles $A$ and $B$ of the
torus respectively, where:
\begin{equation}
\begin{array}{cc}
\theta _{x}=\pi _{x}-\frac{2e}{c}By=-i\hbar \partial _{x}, & \theta _{y}=\pi
_{y}+\frac{2e}{c}Bx=-i\hbar \partial _{y}+\frac{2e}{c}Bx
\end{array}
\label{mgo1}
\end{equation}
and the gauge choice $\overrightarrow{A}\left( x,y\right) =\left(
-By,0\right) $ has been made. They satisfy the relations:
\begin{equation}
\begin{array}{cc}
\left[ \widetilde{{\cal S}},{\cal H}\left( x,y\right) \right] =\left[
\widetilde{{\cal T}},{\cal H}\left( x,y\right) \right] =0, & \widetilde{%
{\cal S}}\widetilde{{\cal T}}=e^{2\pi i\Phi _{ab}/\Phi _{0}}\widetilde{{\cal %
T}}\widetilde{{\cal S}}
\end{array}
,  \label{mgo2}
\end{equation}
where $\Phi _{ab}$\ is the magnetic flux threading the torus surface, and
their action on the wave functions can be defined as:
\begin{equation}
\begin{array}{cc}
{\bf \ }\widetilde{{\cal S}}\varphi \left( x,y\right) =\varphi \left(
x+a,y\right) , & \widetilde{{\cal T}}\varphi \left( x,y\right) =e^{2\pi
iBbx/\Phi _{0}}\varphi \left( x,y+b\right)
\end{array}
.  \label{mgo3}
\end{equation}
Now for $\Phi _{ab}=M\Phi _{0}$ (i.e. when the magnetic flux $\Phi
_{ab}$\ is an integer number of flux quanta $\Phi
_{0}=\frac{hc}{2e}$) the condition $\left[ \widetilde{{\cal
S}},\widetilde{{\cal T}}\right] =0$ holds and we
can simultaneously diagonalize the operators $H\left( x,y\right) $, $%
\widetilde{{\cal S}}$, $\widetilde{{\cal T}}$.\ By introducing adimensional
coordinates on the torus of the kind $T=\left\{ \omega =x+\tau y:x\in \left[
0,1\right] ,y\in \left[ 0,1\right] \right\} $, eqs. (\ref{mgo3}) can be
rewritten as:
\begin{equation}
\begin{array}{cc}
{\bf \ }\widetilde{{\cal S}}\varphi \left( \omega \right) =\varphi \left(
\omega +1\right) , & \widetilde{{\cal T}}\varphi \left( \omega \right)
=e^{2\pi iMx}\varphi \left( \omega +\tau \right)
\end{array}
.  \label{mgo4}
\end{equation}
One can look for eigenfunctions of the kind $\varphi \left( \omega \right)
=e^{i\pi My^{2}\tau }f\left( \omega \right) $ and define magnetic
translation operators $S_{\alpha },T_{\alpha }$\ acting only on $f\left(
\omega \right) $:
\begin{equation}
\begin{array}{cc}
{\bf \ }{\cal S}_{\alpha }f\left( \omega \right) =f\left( \omega +\alpha
\right) , & {\cal T}_{\alpha }f\left( \omega \right) =e^{i\pi M\left( \alpha
^{2}\tau +2\alpha \omega \right) }f\left( \omega +\alpha \tau \right)
\end{array}
.  \label{mgo5}
\end{equation}
In this way eqs. (\ref{mgo4}) become:
\begin{equation}
\begin{array}{cc}
{\bf \ }\widetilde{{\cal S}}\varphi \left( \omega \right) =e^{i\pi
My^{2}\tau }{\cal S}_{1}f\left( \omega \right) , & \widetilde{{\cal T}}%
\varphi \left( \omega \right) =e^{i\pi My^{2}\tau }{\cal T}_{1}f\left(
\omega \right)
\end{array}
.  \label{mgo6}
\end{equation}
Going back to the system of Cooper pairs, in order to describe a coherent
state on a torus we look for wave functions of the kind:
\begin{equation}
\psi _{a}\left( \omega _{1},...,\omega _{M}\right) =e^{i\pi M\tau
\sum_{i=1}^{M}y_{i}^{2}}f_{a}\left( \omega _{1},...,\omega _{M}\right) ,
\label{mgo7}
\end{equation}
\begin{equation}
f_{a}\left( \omega _{1},...,\omega _{M}\right) =\prod_{i<j=1}^{M}\left[
\frac{\theta _{1}\left( \omega _{ij},\tau \right) }{\theta _{1}^{^{\prime
}}\left( 0,\tau \right) }\right] ^{4}\chi _{a}\left( \omega |\tau \right) ,
\label{mgo8}
\end{equation}
where $\omega =\sum_{i=1}^{M}\omega _{i}$ is the ``center of charge''
variable and the non local functions $\chi _{a}\left( \omega |\tau \right) $
are the characters of our theory (the TM). In fact it can be shown that such
characters are eigenfunctions of the following generalized magnetic
translations operators:
\begin{equation}
\begin{array}{cc}
{\bf \ }{\cal S}_{\alpha }=\prod_{i=1}^{M}{\cal S}_{\alpha /M}^{i}, & {\cal T%
}_{\alpha }=\prod_{i=1}^{M}{\cal T}_{\alpha /M}^{i}
\end{array}
,  \label{mgo9}
\end{equation}
where $S_{\alpha /M}^{i}$ and $T_{\alpha /M}^{i}$ are the magnetic
translation operators for the single Cooper pair. In this sense our
characters represent highly non local functions: all the topological
properties of our system are codified in such functions.

On a pure topological base we expect for the torus a doubling of
the ground state degeneracy, which can be seen at the level of the
conformal blocks (characters) of our TM. Indeed we get for the
periodic (even ladder) case an
untwisted sector, $P-P$\ and $P-A$, described by the four conformal blocks (%
\ref{vac1.})-(\ref{vac4.}), and for the Mobius (odd ladder) case a twisted
sector, $A-P$\ and $A-A$, described by the four conformal blocks (\ref{tw1}%
)-(\ref{tw4.}). Now we extract from the vacua of our theory the two states
of the ``protected''qubit.

Let $A$ be the cycle of the torus which surrounds the hole of the Corbino
disk; in the twisted sector it is composed by the leg $1$ and the leg $2$
through the gluing-point as shown in Fig. 1. As underlined in our previous
publications \cite{noi4}, the ground state wave functions of the twisted and
untwisted sectors of the TM are characterized by different {\it monodromy}
properties along the $A$-cycle. In particular the characters of the
untwisted sector are single-valued functions along the $A$-cycle while the
characters of the twisted sector pick up a common $(-1)$ phase factor along
the $A$-cycle. Such phase factors can be interpreted as Bohm-Aharonov phases
generated while a Cooper pair is taken along the $A$-cycle. The above
observation evidences a strong difference between the two inequivalent
topological even ladder (untwisted sector) and odd ladder (twisted sector)
configurations. Indeed in the odd ladder the ground state wave functions
show a non trivial behavior implying the trapping of a half flux quantum ($%
\frac{1}{2}\left( \frac{hc}{2e}\right) $ ) in the hole of the Corbino disk.
Instead in the even ladder, due to the single-valued ground state wave
functions, only integer numbers of flux quantum can be attached to the hole.

It is worth pointing out the central role played by the ``isospin'' (or
neutral) component of the TM in producing the discussed non trivial {\it %
monodromy} properties. To this end let us recall that the TM is a $c=2$ CFT,
composed by a $c=1$ {\it charged} and a $c=1$ {\it isospin }CFT components,
as it is well evidenced by the character decompositions in the Appendix. Now
the transport of a Cooper pair along the $A$-cycle is implemented by\ a
simultaneous and identical translation $\Delta w_{c}=\Delta w_{n}=2$ of the
charged\ and the isospin\ variables. The ``charged'' characters have trivial
monodromy with respect to this transformation, being:
\begin{equation}
K_{l}(w_{c}+2|\tau )=K_{l}(w_{c}|\tau ),\,\ \ l=0,..,3,
\end{equation}
while the ``isospin'' contribution is the one responsible for the non
trivial monodromy of the complete ground state wave functions:
\begin{equation}
\chi _{0,\frac{1}{2}}(2|\tau )=\chi _{0,\frac{1}{2}}(0|\tau )\,,\qquad \chi
_{\frac{1}{16}}(2|\tau )=(-1)\chi _{\frac{1}{16}}(0|\tau )  \label{c18}
\end{equation}
and the same is true for the characters $\bar{\chi}_{\beta }$. Let us notice
that the change in sign in the last relation of eq. (\ref{c18}) shows the
presence in the spectrum of excitations carrying fractionalized charge
quanta. More precisely the presence in the isospin component of one
twist-field (with conformal dimension $\Delta =1/16$) characterizes all the
conformal blocks of the twisted sector and accounts for the trapping of a
half flux quantum in the hole of the closed JJL. At this point it is crucial
to observe that in order to create such a fractionally charged excitation in
the ground state a finite energy must be provided, so assuring the presence
of a finite gap separating the ground state from the excited ones (that is
in complete analogy with the presence of a gap separating the ground state
from higher energy states in the Laughlin Hall fluid \cite{hall1}).

We are now in the position to address the study of the stability and
transformation properties of the ground state wave functions when a magnetic
flux change takes place through the central hole of the closed JJL. The
above analysis shows that at the level of the wave functions it has the
effect to change the monodromy along the $A$-cycle due to the corresponding
change in the Bohm-Aharonov phase. Such a modification can be implemented on
the center of charge component of the wave function, i.e. the characters,
with a well defined transformation. In the case of the {\it charged}
component this analysis has been brought out already in \cite{Napoli-91-92}
in the physical contest of the quantum Hall effect. Let us adapt here the
results for the {\it charged} component of our TM.

On a pure physical ground the fact that we are considering a magnetic flux
change, which is on one side {\it integer} in the flux quantum (one flux
quantum change $\pm \frac{hc}{2e}$) and on the other side {\it adiabatic, }%
suggests both that the monodromy properties do not change and that the
system remains in a degenerate ground state. Such a physical picture is in
fact confirmed for the {\it charged} component of our TM; indeed, the flux
change is implemented on the level of {\it charged }characters by the
transformation ${\cal T}_{1/2}^{c}$:
\begin{equation}
{\cal T}_{1/2}^{c}K_{l}\left( w_{c}|\tau \right) \equiv e^{\left( \frac{1}{2}%
\right) ^{2}i\pi \tau +\frac{2i\pi w_{c}}{2}}K_{l}\left( w_{c}+\frac{\tau }{2%
}|\tau \right) =K_{l+1}\left( w_{c}|\tau \right) ,\ l=0,..,3.
\end{equation}
In particular the {\it charged} component wave functions realize a flip
process ($l\rightarrow l+1$) under one magnetic flux quantum change.

However the analysis for the complete TM, with {\it charged} and {\it isospin%
} components, is more involved. In particular the problem of the stability
of the ground state wave functions under the change of one flux quantum in
the central hole has to be clearly brought out. This is mainly due to the
non trivial interplay between {\it charged} and {\it isospin} components
summarized in the so-called $m$-ality parity rule, which characterizes the
gluing condition for the {\it charged} and {\it isospin} excitations (see
Appendix). The main point being the compatibility between such parity rule
and the transformation of the complete characters of TM under the insertion
of a flux quantum through the hole of the closed ladder, which reads as:
\begin{equation}
{\cal T}_{1/2}f(w_{n}|w_{c}|\tau )=\left. e^{2i\pi (\alpha ^{2}\tau +\alpha
(w_{n}+w_{c}))}f(w_{n}+\alpha \tau |w_{c}+\alpha \tau |\tau )\right|
_{\alpha =1/2},
\end{equation}
where $f(w_{n}=0|w_{c}|\tau )$ stays for any character of TM. The full list
of such transformations is presented in Appendix, here we only comment on
the very simple and clear picture which emerges for the stability and
transformations of the ground states of the closed JJL.

The even configuration of the closed JJL (periodic case) is proven to be
unstable under this transformation. Indeed eq. (\ref{t(1/2)-PA}) and (\ref
{t(1/2)-PPa}) show the decoupling of the untwisted $P-A$ sector and of the
state $\tilde{\chi}_{\alpha }^{+}$ of the $P-P$ sector while eq. (\ref
{t(1/2)-PPbc}) shows that the state $\tilde{\chi}_{\beta }^{+}$ of $P-P$
sector gets excited by this transformation.

For the odd configuration of the closed JJL (Mobius case), eq. (\ref
{t(1/2)-AA}) shows that the twisted $A-A$ sector decouples. So we are left
only with the $A-P$ sector, with the two ground states flipping one into the
other under an adiabatic flux change of $\pm \frac{hc}{2e}$ through the
central hole, as it can be seen from eq. (\ref{t(1/2)-AP}).

Summarizing, between the two inequivalent configurations for the closed JJL,
corresponding to {\it even} and {\it odd} ladder, just the {\it odd} one is
proven to be stable under an adiabatic flux change of $\pm \frac{hc}{2e}$
through the central hole. Then in terms of the ground states {\it center of
charge }wave functions (characters), we can make the following
identifications:
\begin{equation}
\left| 0\right\rangle \backsim \chi _{\left( 0\right) }^{+}(0|w_{c}|\tau
),~~~~~\left| 1\right\rangle \backsim \chi _{\left( 1\right)
}^{+}(0|w_{c}|\tau ).  \label{logical1}
\end{equation}
Then $\left| 0\right\rangle $ and $\left| 1\right\rangle $\ are the two
ground states of the odd closed JJL characterized by the {\it size}
invariant sum over all plaquettes $\sum_{p}\chi _{p}$ respectively equal to $%
-1$ and $+1$.

Based on the above consideration we are ready to propose the odd closed JJL
as our protected qubit. In fact the two ground states $\left| 0\right\rangle
$ and $\left| 1\right\rangle $\ work as the two logical states of the qubit
and the required one qubit operations:
\[
\left| 0\right\rangle \rightarrow \left| 1\right\rangle ,~~~~~\left|
1\right\rangle \rightarrow \left| 0\right\rangle ,
\]
are simply implemented by insertion of a flux quantum ($\pm \frac{hc}{2e}$)
through the central hole.

Let us now make a comment on the stability of such qubit device
under local perturbations. Local perturbations can be viewed in
such a context as creating a finite energy excitation above the
ground states in the form of double kinks. A double kink can be
produced from the ground state by exchanging the chirality of two
nearest neighbor plaquettes in the ladder
and, as such, it is local and it leaves invariant the chirality sum $%
\sum_{p}\chi _{p}$ over all plaquettes and so the characterization
of the two logical states. Furthermore, since a double kink can be
described by the presence of two elementary half flux quanta of
opposite sign ($\pm \frac{1}{2}\frac{hc}{2e}$) localized in
between the pairs of plaquettes with the same chirality, it
doesn't produce any flux change in the central hole. In this way
the excited logical state wave function shows the same monodromy
properties along the $A$-cycle as the corresponding ground state
one and, in particular, satisfies the same transformation rules
under an adiabatic flux change of $\pm \frac{hc}{2e}$ through the
central hole. Summarizing, the characterization of the two logical
states and their flipping processes are left unchanged under local
perturbations, which produce a finite energy excitation above the
ground state.

The minimal configuration for such a protected qubit is represented in Fig.
2 by a closed fully frustrated JJL with $N=3$ plaquettes, 3 being the
minimum odd number of plaquettes needed in order to fulfill all the above
requests.
\begin{figure}[tbp]
\includegraphics[height=.3\textheight]{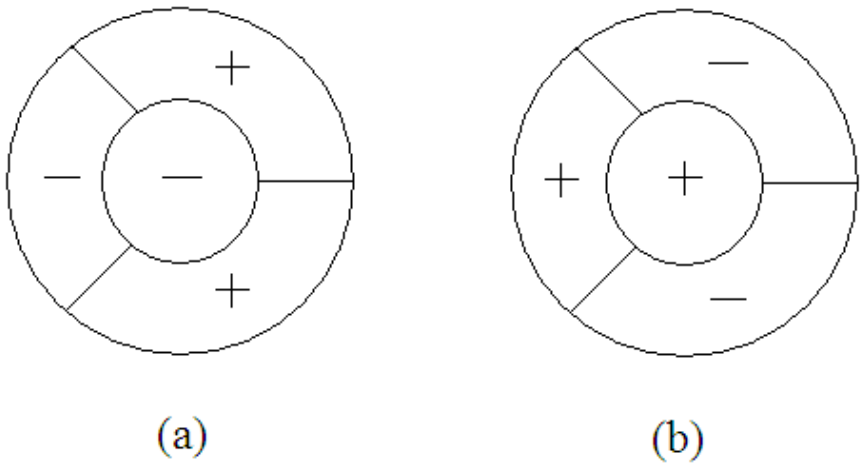}
\caption{The two logical states of the JJL qubit: a) for the state $\left|
1\right\rangle $, b) for the state $\left| 0\right\rangle .$}
\label{figure4}
\end{figure}

Now it should be possible to construct symmetric ($s$) and antisym\-me\-tric
($a$) linear combinations of such degenerate ground states and then to
control their amplitude and relative phase: such operations are needed in
order to prepare the qubit in a definite state and to manipulate it \cite
{qcrew1}. In order to realize the logical $NOT$ we must perform an adiabatic
change of local magnetic fields that drags one vortex across the system,
i.e. a flux quantum through the $A$-cycle of the torus, and flips the state
of the system, so lifting the degeneracy \cite{ioffe}.

Josephson junction ladders with annular geometry have been fabricated within
the trilayer $Nb/Al-AlO_{x}/Nb$ technology and experimentally investigated
\cite{ustinov}, but in such a case the application of an external transverse
magnetic field is needed in order to fulfill the requirement of full
frustration and that could be another source of decoherence. It is now
possible to avoid such a problem by realizing arrays with a built-in
frustration. In fact high-$T_{c}$ Josephson junction arrays have been
recently proposed \cite{rogalla}, which support degenerate spontaneous
current states in zero magnetic field due to the presence of plaquettes
containing an odd number of $\pi $-junctions \cite{pi}. Such unconventional
junctions can be realized because of the $d$-wave symmetry of high-$T_{c}$
superconductors \cite{kirtley}, which produces a $\pi $-shift in the phase
of the wave function on one side of the junction. Furthermore $\pi $%
-junctions can be obtained also with
superconducting-ferromagnetic-superconducting junctions (SFS) \cite{sfs}. In
this way it is possible to avoid the external frustration bias but, in any
case, external magnetic fields are needed for control and read-out
operations: in fact our JJL qubit is a flux device \cite{orlando1}.

So in principle an experimental setup for the realization of our protected
qubit can be conceived: the JJL, arranged in a ring geometry (Corbino disk)
with an odd number of plaquettes along the inner hole, should be equipped
with a coil $P_{bias}$ which can be used to set the system in one of the two
ground states $\left| 0\right\rangle $, $\left| 1\right\rangle $; another
coil $P_{hole}$ is needed in order to control the mixing of the two ground
states and to carry out the flipping. Finally $N$ read-out coils, coupled to
external conventional SQUIDs, are needed in order to read out the state of
the system after the quantum evolution. The whole device has to be embedded
in a superconducting cavity in order to guarantee the stability of the
boundary conditions. In this way a reliable qubit is built up, whose quantum
evolution can be controlled in order to perform all possible single qubit
logical operations. Then such qubits can be antiferromagnetically coupled by
means of suitable superconducting flux transformers (``eight'' coils) which
provide an inductive coupling and whose strength can be controlled within a
wide range of useful values: in this way a quantum register can be realized
and all multi-qubits logical operations can be performed.

Summarizing, a single non-interacting qubit is described by a double well
potential and the external magnetic flux controls the energy difference
between the minima, the symmetric situation being for $\Phi ^{e}=0$. Each
logical state, $\left| 0\right\rangle $ or $\left| 1\right\rangle $, is
represented by a wave function localized in a distinct potential well and
corresponds to distinguishable flux states trapped in the plaquettes of the
ladder with current flowing in opposite directions in alternating
plaquettes. When the energy difference $\varepsilon $ of the minima of the
two different wells is small with respect to the oscillation frequency $%
\omega $ around the minima, $\varepsilon \ll \omega $, these states become
coupled and the wave functions spread over both the wells, the coupling
being maximum in resonance conditions ($\varepsilon =0$), while the energy
eigenstates tend to be localized in one of the well away from resonance as $%
\varepsilon $ is increasing. The coupling of the states can be described by
a tunneling amplitude $\Delta \left( \varepsilon \right) $ and the effective
Hamiltonian of any qubit reduces to the regular two-state form in the basis
of these logical states:
\begin{equation}
H_{eff}=\frac{1}{2}\left[ \varepsilon \left( \left| 0\right\rangle
\left\langle 0\right| -\left| 1\right\rangle \left\langle 1\right| \right)
-\Delta \left( \left| 0\right\rangle \left\langle 1\right| +\left|
1\right\rangle \left\langle 0\right| \right) \right] =\frac{1}{2}\left(
\varepsilon \sigma ^{z}-\Delta \sigma ^{x}\right) ,  \label{qubith}
\end{equation}
where $\sigma ^{x}$, $\sigma ^{z}$ are the Pauli spin matrices.

The diagonal elements of $H_{eff}$ can be easily controlled by an external
magnetic field in the $z$-direction producing an external flux $\Phi ^{e}$
while the off-diagonal elements are related to the tunneling amplitude and
thus are controlled by the adiabatic change of the magnetic flux in the
central hole. The general state vector of such a qubit is the linear
combination of the basis states:
\begin{equation}
\left| \Psi \right\rangle =\alpha \left| 0\right\rangle +\beta \left|
1\right\rangle =\left(
\begin{array}{c}
\alpha \\
\beta
\end{array}
\right) ,  \label{qubitstate}
\end{equation}
so it is described by two complex numbers $\alpha $ and $\beta $.

When the inductive coupling among qubits is turned on, there could be a bias
in, say, the $j$-th qubit even though $\Phi _{j}^{e}=0$ and, as a
consequence, its logical states may be asymmetric. In the approximation in
which every JJL can be considered as a two level system, the system of flux
linked qubits can be described by an effective Hamiltonian of the kind:
\begin{equation}
\overline{H}_{eff}=\sum_{j}\varepsilon _{j}\sigma _{j}^{z}+\sum_{j}\Delta
_{j}\sigma _{j}^{x}+\sum_{\left\langle jk\right\rangle }\Lambda _{kj}\sigma
_{k}^{z}\sigma _{j}^{z}.  \label{registerh}
\end{equation}
In order to control such Hamiltonian, one should be able to modulate the
tunneling amplitude of each qubit as well as to switch on and off the
magnetic coupling between neighbors qubits. The analysis of multi-qubit
logical operations will be the subject of a future publication.

In conclusion in this letter we have presented a simple collective
description of a fully frustrated ladder of Josephson junctions arranged in
a non trivial geometry, with a macroscopic half flux quantum trapped in the
hole. The powerful tools of the CFT have been used to evince the topology,
the stability and the transformation properties of the system. In particular
it has been shown how such features can be exploited for the realization of
a ``protected''\ qubit: a simple device has been proposed and its operation
mode has been briefly sketched.

\section*{Acknowledgments}

We thank P. Minnhagen and K. Schoutens for many enlightening discussions and
for a critical reading of the manuscript. G. N. would like also to thank N.
Kitanine for interesting discussions on a related lattice quantum integrable
model \cite{PhM,Kolya}. G. N. is supported by the ANR programm MIB-05
JC05-52749.

\section*{Appendix: reminder of TM for the JJL}

Here we briefly summarize the main results of our theory, the TM, for the
fully frustrated JJL \cite{noi3}\cite{noi4}. We first construct the bosonic
theory and show that its energy momentum tensor fully reproduces the
Hamiltonian of eq. (\ref{ha3}) for the JJL. That allows us to describe the
JJL excitations in terms of the primary fields $V_{\alpha }\left( z\right) $%
. Then we show that it is possible to construct the $N-$vertices correlator
for the torus topology in $2D$ (basically by letting the edge to evolve in
``time''\ and to interact with external vertex operators placed at different
points). We assume that a suitable correlator is apt to describe the ground
state wave function of the JJL at $T=0$ temperature and then perform an
analysis of the symmetry properties of its center of charge wave function
(conformal blocks), which emerge in the presence of vortices carrying half
quantum of flux ($\frac{1}{2}\left( \frac{hc}{2e}\right) $).

Let us focus on the $m$-reduction procedure \cite{cgm4} for the special $m=2$
case (see Ref. \cite{cgm2} for the general case), since we are interested in
a system with a $Z_{2}$ symmetry and choose the ``bosonic''\ theory \cite
{noi3}\cite{noi4}, which well adapts to the description of a system with
Cooper pairs of electric charge $2e$ in the presence of a topological defect
\cite{noi1}, i.e. a fully frustrated JJL. To each of the two legs (edges) of
the ladder we assign a chirality, so making a correspondence between up-down
leg and left-right chirality states.

Let us now write each phase field as the sum $\varphi ^{\left( a\right)
}\left( x\right) =\varphi _{L}^{\left( a\right) }\left( x\right) +\varphi
_{R}^{\left( a\right) }\left( x\right) $\ of left and right moving fields
defined on the half-line because of the topological defect located in $x=0$.
Then let us define for each leg the two chiral fields $\varphi
_{e,o}^{\left( a\right) }\left( x\right) =\varphi _{L}^{\left( a\right)
}\left( x\right) \pm \varphi _{R}^{\left( a\right) }\left( -x\right) $, each
defined on the whole $x-$axis \cite{boso}. In such a framework the dual
fields $\varphi _{o}^{\left( a\right) }\left( x\right) $\ are fully
decoupled because the corresponding boundary interaction term in the
Hamiltonian does not involve them \cite{affleck}; they are involved in the
definition of the conjugate momenta $\Pi _{\left( a\right) }=\left( \partial
_{x}\varphi _{o}^{\left( a\right) }\right) =\left( \frac{\partial }{\partial
\varphi _{e}^{\left( a\right) }}\right) $\ present in the quantum
Hamiltonian. Performing the change of variables $\varphi _{e}^{\left(
1\right) }=X+\phi $, $\varphi _{e}^{\left( 2\right) }=X-\phi $\ ($\varphi
_{o}^{\left( 1\right) }=\overline{X}+\overline{\phi }$, $\varphi
_{o}^{\left( 2\right) }=\overline{X}-\overline{\phi }$\ for the dual ones)
we get the quantum Hamiltonian (\ref{ha3}) but now all the fields are chiral
ones. Finally let us identify in the continuum such chiral phase fields $%
\varphi _{e}^{\left( a\right) }$, $a=1,2$, each defined on the corresponding
leg, with the two chiral fields $Q^{\left( a\right) }$, $a=1,2$\ of the TM
with central charge $c=2$.

As a result of the $2$-reduction procedure \cite{cgm2}\cite{cgm4} we get a $%
c=2$ orbifold CFT, the TM, whose fields have well defined transformation
properties under the discrete $Z_{2}$ (twist) group, which is a symmetry of
the TM. Its primary fields content can be expressed in terms of a $Z_{2}$%
-invariant scalar field $X(z)$, given by
\begin{equation}
X(z)=\frac{1}{2}\left( Q^{(1)}(z)+Q^{(2)}(z)\right) ,  \label{X}
\end{equation}
describing the continuous phase sector of the theory, and a twisted field
\begin{equation}
\phi (z)=\frac{1}{2}\left( Q^{(1)}(z)-Q^{(2)}(z)\right) ,  \label{phi}
\end{equation}
which satisfies the twisted boundary conditions $\phi (e^{i\pi }z)=-\phi (z)$
\cite{cgm2}. More explicitly such a field can be written in terms of the
left and right moving components $\varphi _{L}^{\left( 1\right) }$, $\varphi
_{R}^{\left( 2\right) }$ as we stated above; then the Mobius boundary
conditions given in eq. (\ref{blr}) are described by the boundary conditions
for $\phi $. This will be more evident for closed geometries, i.e. for the
torus case, where the magnetic impurity gives rise to a line defect in the
bulk, so allowing us to resort to the folding procedure and introduce
boundary states \cite{noi1}\cite{noi2}. Such a procedure is used in the
literature to map a problem with a defect line (as a bulk property) into a
boundary one, where the defect line appears as a boundary state of a theory
which is not anymore chiral and its fields are defined in a reduced region
which is one half of the original one. Our approach, the TM, is a chiral
description of that, where the chiral $\phi $\ field defined in ($-L/2$, $%
L/2)$ describes both the left moving component and the right moving one
defined in ($-L/2$, $\ 0$), ($0$, $L/2$) respectively, in the folded
description \cite{noi1}\cite{noi2}. Furthermore to make a connection with
the TM we consider more general gluing conditions:
\[
\phi _{L}(x=0)=\mp \phi _{R}(x=0),
\]
the $-$($+$) sign staying for the twisted (untwisted) sector. We are then
allowed to use the boundary states given in \cite{Affleck} for the $c=1$
orbifold at the Ising$^{2}$ radius. The $X$ field, which is even under the
folding procedure, does not suffer any change in boundary conditions \cite
{noi1} while condition (\ref{blr}) is naturally satisfied by the twisted
field $\phi \left( z\right) $. So topological order can be discussed
referring to the characters with the implicit relation to the different
boundary states (BS) present in the system \cite{noi1}. These BS should be
associated to different kinds of linear defects compatible with conformal
invariance.

The fields in eqs. (\ref{X})-(\ref{phi}) coincide with the ones introduced
in eq. (\ref{ha3}). In fact the energy momentum tensor for such fields fully
reproduces the second quantized Hamiltonian of eq. (\ref{ha3}). The whole TM
theory decomposes into a tensor product of two CFTs, a twisted invariant one
with $c=\frac{3}{2}$ and the remaining $c=\frac{1}{2}$ one realized by a
Majorana fermion in the twisted sector. In the $c=\frac{3}{2}$ sub-theory
the primary fields are composite vertex operators $V\left( z\right)
=U_{X}^{\alpha _{l}}\left( z\right) \psi \left( z\right) $ or $V_{qh}\left(
z\right) =U_{X}^{\alpha _{l}}\left( z\right) \sigma \left( z\right) $, where
\begin{equation}
U_{X}^{\alpha _{l}}\left( z\right) =\frac{1}{\sqrt{z}}:e^{i\alpha _{l}X(z)}:
\label{char}
\end{equation}
is the vertex of the continuous\ sector with $\alpha _{l}=\frac{l}{2}$, $%
l=1,...,4$ for the $SU(2)$ Cooper pairing symmetry used here. Regarding the
other\ component, the highest weight state in the isospin sector, it can be
classified by the two chiral operators:
\begin{equation}
\psi \left( z\right) =\frac{1}{2\sqrt{z}}\left( :e^{i\sqrt{2}\phi \left(
z\right) }:+:e^{i\sqrt{2}\phi \left( -z\right) }:\right) ,~~~~~~\overline{%
\psi }\left( z\right) =\frac{1}{2\sqrt{z}}\left( :e^{i\sqrt{2}\phi \left(
z\right) }:-:e^{i\sqrt{2}\phi \left( -z\right) }:\right) ;  \label{neu1}
\end{equation}
which correspond to two $c=\frac{1}{2}$ Majorana fermions with Ramond
(invariant under the $Z_{2}$ twist) or Neveu-Schwartz ($Z_{2}$ twisted)
boundary conditions \cite{cgm2}\cite{cgm4} in a fermionized version of the
theory. The Ramond fields are the degrees of freedom which survive after the
tunnelling and the parity symmetry, which exchanges the two Ising fermions,
is broken. Besides the fields appearing in eq. (\ref{neu1}), there are the $%
\sigma \left( z\right) $ fields, also called the twist fields, which appear
in the quasi-hole primary fields $V_{qh}\left( z\right) $. The twist fields
have non local properties and decide also for the non trivial properties of
the vacuum state, which in fact can be twisted or not in our formalism.

Starting from the primary fields $V_{\alpha }\left( z\right) $ we can now
construct the non perturbative ground state wave function of the JJL system
for the torus topology. It turns out that by construction it results as a
coherent superposition of gaussian states with all the non trivial global
properties of the order parameter.{\bf \ }In fact by using standard
conformal field theory techniques it is now possible to generate the torus
topology, starting from the edge theory, just defined above. That is
realized by evaluating the $N$-vertices correlator
\begin{equation}
\left\langle n\right| V_{\alpha }\left( z_{1}\right) \ldots V_{\alpha
}\left( z_{N}\right) e^{2\pi i\tau L_{0}}\left| n\right\rangle ,
\end{equation}
where $V_{\alpha }\left( z_{i}\right) $ is the generic primary field
representing the excitation at $z_{i}$, $L_{0}$ is the Virasoro generator
for dilatations and $\tau $ the proper time. The neutrality condition $\sum
\alpha =0$ must be satisfied and the sum over the complete set of states $%
\left| n\right\rangle $ is indicating that a trace must be taken. It is very
illuminating for the non expert reader to pictorially represent the above
operation in terms of an edge state (that is a primary state defined at a
given $\tau $) which propagates interacting with external fields at $%
z_{1}\ldots z_{N}$ and finally getting back to itself. In such a way a $2D$
surface is generated with the torus topology. It is interesting to observe
that such a procedure is equivalent to the coherent insertion of correlated
relevant vortices (as provided by the CFT description) at positions $%
z_{1}\ldots z_{N}$, as they appear in the non perturbative ground state of
the physical JJL system.{\bf \ }From such a picture it is evident then how
the degeneracy of the non perturbative ground state is closely related to
the number of primary states. Furthermore, since in this letter we are
interested in the understanding of the topological properties of the system,
we can consider only the center of charge contribution in the above
correlator, so neglecting its short distances properties. To such an extent
the one-point functions are extensively reported in the following.

On the torus \cite{cgm4} the TM primary fields are described in terms of the
conformal blocks of the $Z_{2}$-invariant $c=\frac{3}{2}$ sub-theory and of
the non invariant $c=\frac{1}{2}$ Ising model, so reflecting the
decomposition on the plane above outlined. The characters $\bar{\chi}%
_{0}(0|\tau )$, $\bar{\chi}_{\frac{1}{2}}(0|\tau )$, $\bar{\chi}_{\frac{1}{16%
}}(0|\tau )$ express the primary fields content of the Ising model \cite{cft}
with Neveu-Schwartz ($Z_{2}$ twisted) boundary conditions \cite{cgm4}, while
\begin{eqnarray}
\chi _{(0)}^{c=3/2}(0|w_{c}|\tau ) &=&\chi _{0}(0|\tau )K_{0}(w_{c}|\tau
)+\chi _{\frac{1}{2}}(0|\tau )K_{2}(w_{c}|\tau )\,,  \label{mr1} \\
\chi _{(1)}^{c=3/2}(0|w_{c}|\tau ) &=&\chi _{\frac{1}{16}}(0|\tau )\left(
K_{1}(w_{c}|\tau )+K_{3}(w_{c}|\tau )\right) ,  \label{mr2} \\
\chi _{(2)}^{c=3/2}(0|w_{c}|\tau ) &=&\chi _{\frac{1}{2}}(0|\tau
)K_{0}(w_{c}|\tau )+\chi _{0}(0|\tau )K_{2}(w_{c}|\tau )  \label{mr3}
\end{eqnarray}
represent those of the $Z_{2}$-invariant $c=\frac{3}{2}$ CFT. They are given
in terms of a ``charged''\ $K_{\alpha }(w_{c}|\tau )$ contribution:
\begin{equation}
K_{2l+i}(w|\tau )=\frac{1}{\eta \left( \tau \right) }\;\Theta \left[
\begin{array}{c}
\frac{2l+i}{4} \\[6pt]
0
\end{array}
\right] (2w|4\tau )\,,\qquad \text{with }l=0,1\text{ and }i=0,1\,,
\label{chp}
\end{equation}
and a ``isospin''\ one $\chi _{\beta }(0|\tau )$, (the conformal blocks of
the Ising Model), where $w_{c}=\dfrac{1}{2\pi i}\,\ln z_{c}$ is the torus
variable of the ``charged''\ component while the corresponding argument of
the isospin block is $w_{n}=0$ everywhere.

If we now turn to the whole $c=2$ theory, the characters of the twisted
sector are given by:
\begin{eqnarray}
\chi _{(0)}^{+}(0|w_{c}|\tau ) &=&\bar{\chi}_{\frac{1}{16}}(0|\tau )\left(
\chi _{0}+\chi _{\frac{1}{2}}\right) (0|\tau )\left( K_{0}+K_{2}\right)
(w_{c}|\tau ),  \label{tw1} \\
\chi _{(1)}^{+}(0|w_{c}|\tau ) &=&\chi _{\frac{1}{16}}(0|\tau )\left( \bar{%
\chi}_{0}+\bar{\chi}_{\frac{1}{2}}\right) (0|\tau )\left( K_{1}+K_{3}\right)
(w_{c}|\tau ),  \label{tw2}
\end{eqnarray}
for the $A-P$ sector and by:
\begin{eqnarray}
\chi _{(0)}^{-}(0|w_{c}|\tau ) &=&\bar{\chi}_{\frac{1}{16}}(0|\tau )\left(
\chi _{0}-\chi _{\frac{1}{2}}\right) (0|\tau )\left( K_{0}-K_{2}\right)
(w_{c}|\tau ),  \label{tw3.} \\
\chi _{(1)}^{-}(0|w_{c}|\tau ) &=&\chi _{\frac{1}{16}}(0|\tau )\left( \bar{%
\chi}_{0}-\bar{\chi}_{\frac{1}{2}}\right) (0|\tau )\left( K_{1}+K_{3}\right)
(w_{c}|\tau ),  \label{tw4.}
\end{eqnarray}
for the $A-A$ one. Furthermore the characters of the untwisted sector are
\cite{cgm4}:
\begin{align}
\tilde{\chi}_{(0)}^{-}(0|w_{c}|\tau )& =\left( \bar{\chi}_{0}\chi _{0}-\bar{%
\chi}_{\frac{1}{2}}\chi _{\frac{1}{2}}\right) (0|\tau )K_{0}(w_{c}|\tau
)+\left( \bar{\chi}_{0}\chi _{\frac{1}{2}}-\bar{\chi}_{\frac{1}{2}}\chi
_{0}\right) (0|\tau )K_{2}\,(w_{c}|\tau ),  \label{vac1.} \\
\tilde{\chi}_{(1)}^{-}(0|w_{c}|\tau )& =\left( \bar{\chi}_{0}\chi _{\frac{1}{%
2}}-\bar{\chi}_{\frac{1}{2}}\chi _{0}\right) (0|\tau )K_{0}(w_{c}|\tau
)+\left( \bar{\chi}_{0}\chi _{0}-\bar{\chi}_{\frac{1}{2}}\chi _{\frac{1}{2}%
}\right) (0|\tau )K_{2}\,(w_{c}|\tau ),
\end{align}
for the $P-A$ sector while for the $P-P$ sector we have:
\begin{align}
\tilde{\chi}_{\alpha }^{+}(0|w_{c}|\tau )& =\frac{1}{2}\left( \bar{\chi}_{0}-%
\bar{\chi}_{\frac{1}{2}}\right) (0|\tau )\left( \chi _{0}-\chi _{\frac{1}{2}%
}\right) (0|\tau )(K_{0}-K_{2})(w_{c}|\tau )\,, \\
\tilde{\chi}_{\beta }^{+}(0|w_{c}|\tau )& =\frac{1}{2}\left( \bar{\chi}_{0}+%
\bar{\chi}_{\frac{1}{2}}\right) (0|\tau )\left( \chi _{0}+\chi _{\frac{1}{2}%
}\right) (0|\tau )(K_{0}+K_{2})(w_{c}|\tau ),  \label{vac4.}
\end{align}
and
\begin{equation}
\tilde{\chi}_{\gamma }^{+}(0|w_{c}|\tau )=\bar{\chi}_{\frac{1}{16}}(0|\tau
)\chi _{\frac{1}{16}}(0|\tau )\left( K_{1}+K_{3}\right) (w_{c}|\tau ).
\end{equation}
Let us comment that the above factorization expresses the parity selection
rule ($m$-ality), which gives a gluing condition for the ``charged''\ and
``isospin''\ excitations.

It is worth underlining that in the $P-P$ sector, unlike for the other
sectors, modular invariance constraint requires the presence of three
different characters. The {\it isospin} operator content of the character $%
\tilde{\chi}_{\gamma }^{+}(0|w_{c}|\tau )$ clearly evidences its peculiarity
with respect to the other states of the periodic (even ladder) case. Indeed
it is characterized by two twist fields ($\Delta =1/16$) in the {\it isospin}
components. The occurrence of the {\it double} twist in the state described
by $\tilde{\chi}_{\gamma }^{+}(0|w_{c}|\tau )$ is simply the reason why such
a state is a periodic state. Indeed, being an {\it isospin} twist field the
representation in the continuum limit of a magnetic impurity (a half flux
quantum trapping\ or equivalently a kink), the double twist corresponds to a
double half flux quantum trapping, i.e. one flux quantum, typical of the
periodic configuration.

The above analysis would suggest that the $P-P$ state described by $\tilde{%
\chi}_{\gamma }^{+}(0|w_{c}|\tau )$ embeds in the continuum limit a
kink-antikink excitation, i.e. it represents an excited state in the $P-P$
sector. In this way, as it happens for all the other sectors, the $P-P$
sector is left with just two degenerate ground states ( $\tilde{\chi}%
_{\alpha }^{+}(0|w_{c}|\tau )$ and $\tilde{\chi}_{\beta }^{+}(0|w_{c}|\tau )$%
) and, as expected on a pure topological base, the ground state degeneracy
in the torus topology is the double of that of the disk.

Let us now present the full list of character transformations under the
insertion of a magnetic flux quantum through the hole of the closed ladder.

In the even closed JJ ladder configuration, we have that the two ground
state wave functions of the $P-A$ sector decouple, being
\begin{equation}
{\cal T}_{1/2}\tilde{\chi}_{(0)}^{-}(0|w_{c}|\tau )=0\,,\text{ \ }{\cal T}%
_{1/2}\tilde{\chi}_{(1)}^{-}(0|w_{c}|\tau )=0.  \label{t(1/2)-PA}
\end{equation}
Concerning the $P-P$ sector, we have:
\begin{equation}
{\cal T}_{1/2}\tilde{\chi}_{\alpha }^{+}(0|w_{c}|\tau )=0  \label{t(1/2)-PPa}
\end{equation}
and
\begin{equation}
{\cal T}_{1/2}\tilde{\chi}_{\beta }^{+}(0|w_{c}|\tau )=\tilde{\chi}_{\gamma
}^{+}(0|w_{c}|\tau )\,\text{ \ \ \ \ }(\ {\cal T}_{1/2}\tilde{\chi}_{\gamma
}^{+}(0|w_{c}|\tau )=\tilde{\chi}_{\beta }^{+}(0|w_{c}|\tau )\,\text{\ }).
\label{t(1/2)-PPbc}
\end{equation}
Such transformations show the instability of the $P-P$ sector under the
insertion of a flux quantum through the hole of the closed ladder. More
precisely the state $\tilde{\chi}_{\alpha }^{+}(0|w_{c}|\tau )$ decouples
while the state $\tilde{\chi}_{\beta }^{+}(0|w_{c}|\tau )$ gets excited to
the state with a kink-antikink configuration $\tilde{\chi}_{\gamma
}^{+}(0|w_{c}|\tau )$.

Furthermore in the odd closed JJ ladder configuration, we have that the two
ground state wave functions of the $A-A$ sector decouple, being
\begin{equation}
{\cal T}_{1/2}\chi _{(0)}^{-}(0|w_{c}|\tau )=0\,,\text{ \ }{\cal T}%
_{1/2}\chi _{(1)}^{-}(0|w_{c}|\tau )=0.  \label{t(1/2)-AA}
\end{equation}
Concerning the $A-P$ sector, we have that the two ground state wave
functions transform as:
\begin{equation}
{\cal T}_{1/2}\chi _{(0)}^{+}(0|w_{c}|\tau )=\chi _{(1)}^{+}(0|w_{c}|\tau
)\,,\text{ \ }{\cal T}_{1/2}\chi _{(1)}^{+}(0|w_{c}|\tau )=\chi
_{(0)}^{+}(0|w_{c}|\tau )\,.  \label{t(1/2)-AP}
\end{equation}

Concluding, the full set of transformations, here presented, allows to claim
the following simple and clear picture: {\it the odd closed JJL
configuration is the only one which is stable under the insertion of a
magnetic flux quantum through the central hole; moreover, in such odd JJL
configuration such a magnetic flux insertion simply implements the flipping
process between the two degenerate ground states }$\left| 0\right\rangle $%
{\it \ and }$\left| 1\right\rangle ${\it .}

\end{document}